\documentclass[preprint,5p]{elsarticle}

\usepackage{amssymb,amsmath}
\usepackage{graphicx}
\usepackage{dcolumn}
\usepackage{bm}

\journal{Physics Letters B}

\begin{document}
\begin{frontmatter}

\title{A direct measurement of the $^{17}$O($\alpha,\gamma$)$^{21}$Ne reaction in inverse kinematics and its impact on heavy element production }%

\author[1]{M.P. Taggart \fnref{fn1}}
\author[1]{C. Akers \fnref{fn2}}
\author[1,13,14]{A.M. Laird\corref{cor1}}
\ead{alison.laird@york.ac.uk}
\author[2]{U. Hager}
\author[2]{C. Ruiz}
 \author[2]{D. A.~Hutcheon}
  \author[1]{M. A. Bentley}
 \author[1]{J.~R.~Brown}
 \author[2]{L. Buchmann} 
\author[3]{A.A. Chen}  
 \author[3]{J. Chen} 
 \author[1]{K. A. Chipps \fnref{fn3}}

 \author[4]{A. Choplin \fnref{fn4}}
  \author[5]{J. M.~D'Auria}
 \author[2,5]{B.~Davids}
  \author[2]{C.~Davis}
  \author[1]{C. Aa. Diget}
 \author[6]{L. Erikson}
  \author[2]{J.~Fallis}
   \author[1]{S. P. Fox}
  \author[7]{U. Frischknecht} 
   \author[1]{B.~R.~Fulton}  
 \author[2]{N.~Galinski}  
 \author[6]{U. Greife}
 \author[8,9,13,14]{R. Hirschi}
\author[2]{D.~Howell}     
\author[2]{L.~Martin}     
 \author[10]{D. Mountford}
 \author[10]{A.~St.J.~Murphy}    
  \author[2]{D.~Ottewell} 
               
 \author[11,12,13,14]{M. Pignatari \fnref{fn5}}
               
\author[2]{S.~Reeve}
\author[2]{G. Ruprecht}
\author[2]{S.~Sjue}
\author[2]{L.~Veloce}
\author[1,2]{M.~Williams}

\cortext[cor1]{Corresponding author}
\fntext[fn1]{Present address: Department of Physics, University of Surrey, U.K.}
\fntext[fn2]{Present address: Rare Isotope Science Project, Institute for Basic Science, Daejeon 34047, Republic of Korea}
               
 \fntext[fn3]{Present address: Physics Division, Oak Ridge National Laboratory, Oak Ridge TN 37831 USA}       
         
 \fntext[fn4]{Present address: Department of Physics, Faculty of Science and Engineering, Konan University, 8-9-1 Okamoto, Kobe, Hyogo 658-8501, Japan }       
                              
\fntext[fn5]{Present address: E.~A.~Milne Centre for Astrophysics, Department of Physics and Mathematics, University of Hull, HU6 7RX, United Kingdom}

\address[1] {Department of Physics, University of York, York, YO10 5DD, UK}
\address[2]{TRIUMF, Vancouver, Canada V6T 2A3}
\address[3]{McMaster University, Hamilton, ON, Canada}
\address[4]{Geneva Observatory, University of Geneva, Maillettes 51, CH-1290 Sauverny, Switzerland}
\address[5]{Simon Fraser University, Burnaby, BC, Canada}
\address[6]{Colorado School of Mines, Golden, CO, USA}
\address[7]{Department of Physics, University of Basel, Klingelbergstrasse 82, 4056 Basel, Switzerland}
\address[8]{Astrophysics Group, Lennard-Jones Labs 2.09, Keele University, ST5 5BG, Staffordshire, UK} 
\address[9]{Kavli Institute for the Physics and Mathematics of the Universe (WPI), University of Tokyo, 5-1-5 Kashiwanoha, Kashiwa, 277-8583, Japan}
 \address[10]{SUPA, School of Physics and Astronomy, The University of Edinburgh, Edinburgh, EH9 3FD, UK}
 \address[11]{University of Victoria, Victoria, BC, Canada}
  
 \address[13]{UK Network for Bridging the Disciplines of Galactic Chemical Evolution (BRIDGCE)}

\address[12]{Konkoly Observatory, Research Centre for Astronomy and Earth Sciences, Hungarian Academy of Sciences, Konkoly Thege Miklos ut 15-17, H-1121 Budapest, Hungary}
\address[14]{ NuGrid Collaboration}

 
\date{\today}

\begin{abstract}
During the slow neutron capture process in massive stars, reactions on light elements can both produce and absorb neutrons thereby influencing the final heavy element abundances. At low metallicities, the high neutron capture rate of $^{16}$O can inhibit s-process nucleosynthesis unless the neutrons are recycled via the $^{17}$O($\alpha$,n)$^{20}$Ne reaction. The efficiency of this neutron recycling is determined by competition between the $^{17}$O($\alpha$,n)$^{20}$Ne and $^{17}$O($\alpha,\gamma$)$^{21}$Ne reactions. While some experimental data are available on the former reaction, no data exist for the radiative capture channel at the relevant astrophysical energies.

The $^{17}$O($\alpha,\gamma$)$^{21}$Ne reaction has been studied directly using the DRAGON recoil separator at the TRIUMF Laboratory. The reaction cross section has been determined at energies between 0.6 and 1.6 MeV E$_{cm}$, reaching into the Gamow window for core helium burning for the first time. Resonance strengths for resonances at 0.63, 0.721, 0.81 and 1.122 MeV E$_{cm}$ have been extracted. The experimentally based reaction rate calculated represents a lower limit, but suggests that significant s-process nucleosynthesis occurs in low metallicity massive stars.

\begin{description}
\item[PACS numbers] 26.20.Kn, 25.40.Lw
\end{description}

\end{abstract}

                             
 \end{frontmatter}
                              

\section{Introduction}
Almost all the elements in the Universe heavier than iron are produced by neutron-capture reactions, either 
via the r-process (rapid neutron capture) or the s-process (slow neutron capture). While significant 
uncertainties remain in r-process nucleosynthesis, the s-process is considered generally well understood. Here, the neutron flux is such that the timescales for neutron capture are longer than the associated $\beta$-decays, and so the path of nucleosynthesis lies close to the valley of stability. Most s-process elements between iron and strontium are thought to have been produced in
massive stars, through the weak s-process, and those between strontium and lead via the main s-process in Asymptotic Giant Branch (AGB) stars \cite{sp}. \\
\indent However, abundance ratios (e.g. [Y/Ba]) observed in extremely metal poor stars and in one of the oldest globular clusters in the galactic bulge, NGC 6522, cannot be explained by the main s-process or the r-process. Chiappini {\it et al.} \cite{C2011} show that massive rotating stars at low metallicity can provide an explanation for the unique abundances observed both in the galactic halo and NGC 6522 (see also \cite{Cesc} and \cite{Chop0}).
For such stars, rotation-induced mixing is considered to have a significant impact on nucleosynthesis of light elements, especially at low metallicities \cite{UF,Hirschi}. S-process abundances depend critically on the presence of those light elements which can act as neutron sources and poisons (isotopes which capture neutrons, thus removing them from contributing to s-process production). At low metallicities, the lack of secondary neutron poisons (e.g. $^{14}$N) and the large abundance of primary $^{16}$O results in a high neutron capture rate to $^{17}$O. Thus $^{16}$O could act as a poison if these neutrons are not recycled via the $^{17}$O($\alpha$,n)$^{20}$Ne reaction. This recycling of neutrons is determined by competition between the $^{17}$O($\alpha$,n)$^{20}$Ne and $^{17}$O($\alpha,\gamma$)$^{21}$Ne reactions. However, these
reaction rates are highly uncertain at the relevant energies and the status of $^{16}$O as a neutron poison, and the impact on s-process abundances, is therefore as yet undetermined. \\
\indent There are two theoretical calculations of the  $^{17}$O($\alpha,\gamma$)$^{21}$Ne to $^{17}$O($\alpha$,n)$^{20}$Ne reaction rate ratio. The first, from Caughlan and Fowler (CF88) \cite{CF88}, 
assumes the ratio to be around 0.1 at low energies, dropping to 5 $\times$ 10$^{-4}$ above about 1 MeV.
This assumption is based on experimental data on the $^{18}$O($\alpha,\gamma$)$^{22}$Ne reaction for the higher energies, and on Hauser-Feshbach calculations at lower energies. The second
prediction comes from Descouvemont \cite{Desc}, using the Generator Coordinate Method, and 
suggests the ratio to be of the order of 10$^{-4}$ at all energies. 
This huge disagreement at low energies results in significant
differences in the predicted s-process abundances. Models by Hirschi {\it et al.} \cite{Hirschi} show the impact of the two different predictions on the abundances of the heavy elements.
The variation is particularly marked (up to three orders of magnitude) between strontium and barium.

\begin{figure}[h!]
\centering\includegraphics[scale=0.5,angle=0]{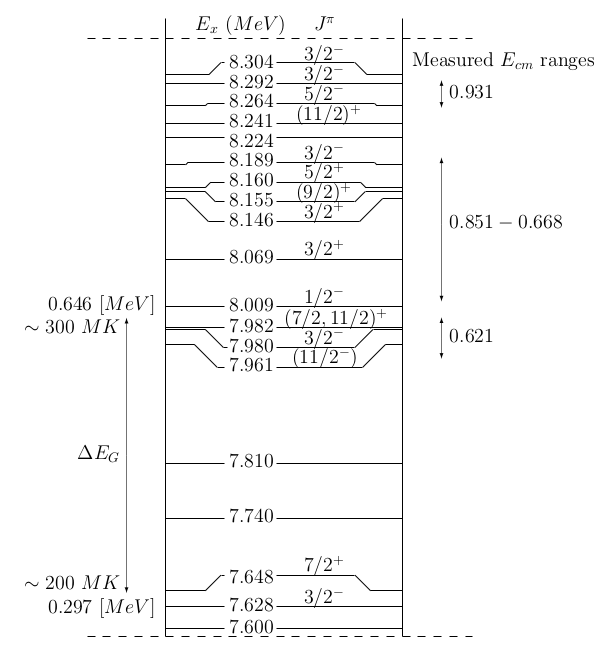}
\caption{\small Part of $^{21}$Ne level scheme. The Gamow window for the $^{17}$O($\alpha,\gamma$)$^{21}$Ne reaction during core helium burning in massive stars is indicated by the bar on the left and the bars on the right show the energy regions covered by the present work.  }
\label{21Nelevels}
\end{figure}

\indent For low metallicity massive stars, s-process nucleosynthesis is thought to occur during two stages of evolution, firstly core helium burning and then later shell carbon burning. The temperature for core helium burning is around 0.2 - 0.3 GK, corresponding to an energy range of interest (Gamow window) between about 0.3 and 0.65 MeV in the centre of mass (E$_{cm}$).
For the onset of carbon shell burning, temperatures are higher at around 0.8 to 1.3 GK, with a Gamow window
between  E$_{cm}$ = 0.7 to 1.8 MeV.  The $^{17}$O($\alpha,\gamma$)$^{21}$Ne reaction Q-value is 7.348 MeV \cite{FS} and the relevant excited states, shown in Figure \ref{21Nelevels}, lie between 7.65 and 8.0 MeV excitation energy (E$_x$) in $^{21}$Ne for core helium burning temperatures. However, the required partial width and spin-parity information for $^{21}$Ne levels in the region of interest is poorly known, preventing reliable calculation of the contribution of individual resonances to the reaction rate.

\indent
Experimental data on the $^{17}$O($\alpha$,n)$^{20}$Ne reaction are available covering the range E$_{cm}$ = 0.56 - 10.1 MeV  \cite{Denker,Bair,Hansen,Best2}, and there is only one published experimental dataset on the $^{17}$O($\alpha,\gamma$)$^{21}$Ne reaction \cite{Best1}.
Traditionally, experimental determinations of such ($\alpha,\gamma$) reaction cross sections have relied on using an intense beam of $\alpha$-particles and the detection of $\gamma$-rays from de-excitation of the products. For the $^{17}$O($\alpha,\gamma$)$^{21}$Ne reaction, however, the high Q-value of the reaction results in the products having high excitation energies where many nuclear states are populated. Clean identification of these states is difficult to extract from the background, particularly at the astrophysically interesting energies where the yield from the reactions of interest is extremely low, typically less than 1 event for every 10$^{12}$ incident $\alpha$-particles. Despite the experimental challenges, measurements using this technique provided the first direct data on the $^{17}$O($\alpha,\gamma$)$^{21}$Ne reaction. Best {\it et al.} \cite{Best1} measured the $^{17}$O($\alpha,\gamma$)$^{21}$Ne reaction by in-beam spectroscopy, using a $^4$He beam on an implanted target.  The measurements spanned E$_{cm}$ between 0.7 and 1.9 MeV but no yield was observed below E$_{cm}$ = 1.1 MeV ($\sim$ 1 GK) except for a strong resonance at E$_{cm}$ = 0.811 MeV, believed to correspond to a state at 8.159(2) MeV. Subsequently Best {\it et al.} \cite{Best2} also studied the competing $^{17}$O($\alpha$,n)$^{21}$Ne reaction across the same energy range. Many resonances were observed and fitted using an R-matrix framework. Finally, using both datasets and estimates for the contribution from lower-lying states, Best et al. \cite{Best2} calculated new reaction rates and concluded that the ($\alpha,\gamma$) channel is strong enough to compete with the ($\alpha$,n) channel leading to less efficient neutron recycling.
However, neither measurement had sufficient sensitivity to provide any experimental data in the energy region relevant to the s-process during the core helium burning stage. \\ 

\section{Experimental details}
\begin{figure}[h!]
\centering\includegraphics[scale=0.35,angle=0]{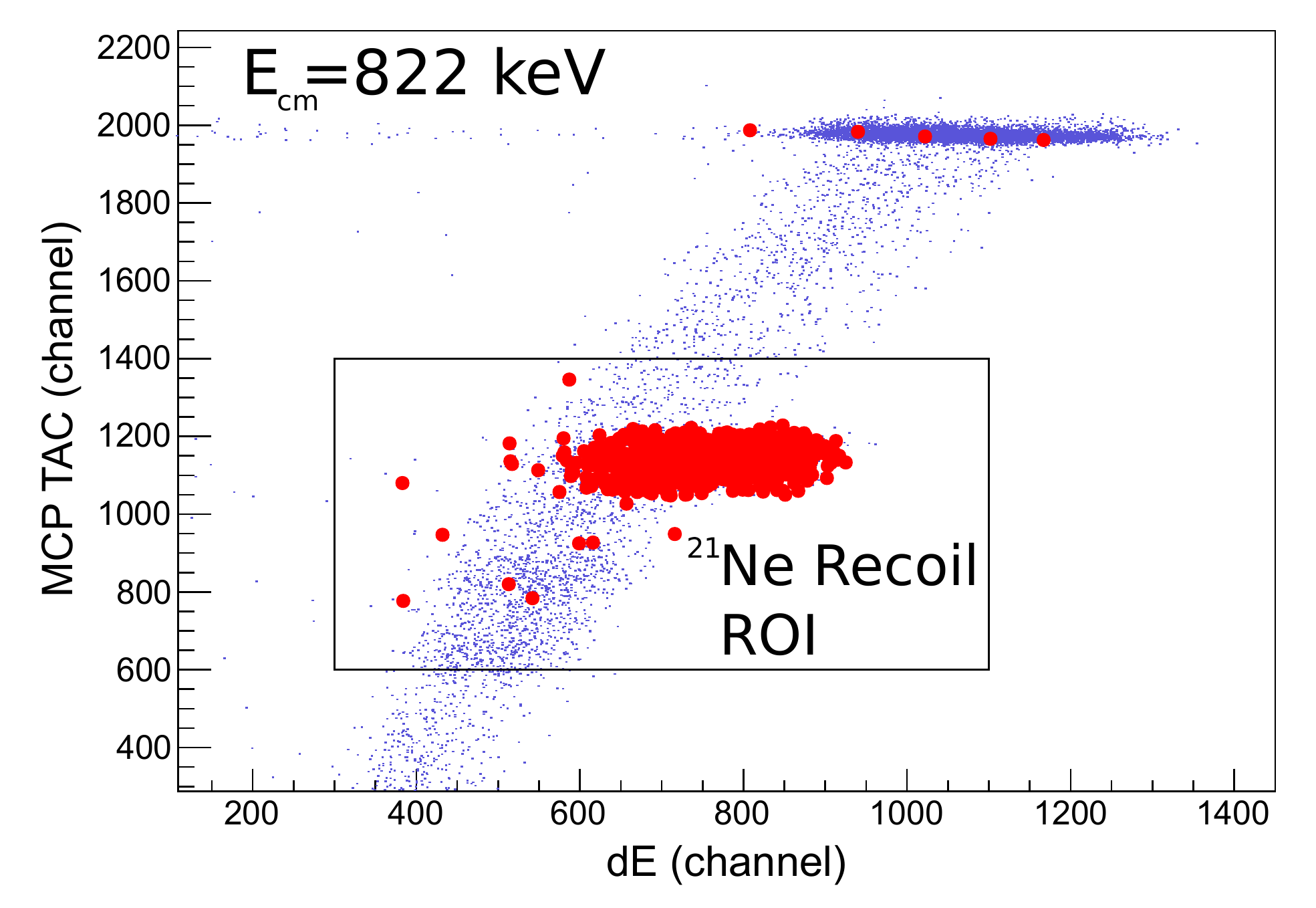}
\caption{\small 
(Color online) MCP local time of flight TAC (time to amplitude converter output) versus dE for data at E$_{cm}$ = 822 keV. Singles events are indicated in blue and coincident data, with a detected gamma-ray energy above 2 MeV, in red. The events in red in the top right of the figure are random coicidences between a gamma-ray and a scattered beam ion.}
\label{etof}
\end{figure}

Here we report on the first 
measurement of the  
$^{17}$O($\alpha,\gamma$)$^{21}$Ne reaction exploiting, instead, a beam of $^{17}$O ions incident on a helium gas target. The $^{21}$Ne recoils from the reaction exited the target (unlike in the above case) with the unreacted beam, allowing their detection in coincidence with prompt gamma rays from their de-excitation. The measurement was performed at the DRAGON recoil separator in the ISAC facility, at the TRIUMF Laboratory, Canada, which is specifically designed to study such radiative capture reactions relevant to nuclear astrophysics. It consists of a windowless recirculating gas target, surrounded by an array of 30 bismuth germanate (BGO)
gamma-ray detectors, and a two-stage electromagnetic recoil separator. Details of the DRAGON separator are given in Hutcheon et al. \cite{Hutch} and Engel et al. \cite{Engel}. \\
\indent The $^{17}$O$^{3+}$ beam with a typical current of 600 enA (corresponding to $\sim$ 1.25 x 10$^{12}$ pps) impinged on the windowless helium gas target. 
DRAGON was configured to transmit 4$^+$ $^{21}$Ne recoils from the $^{17}$O($\alpha,\gamma$)$^{21}$Ne reaction. These recoils were detected at the focal plane by an ionization chamber (IC).
The IC anode consisted of four segments, providing energy loss and residual energy (dE-E) information, and was filled with isobutane at a typical 
pressure of 8 Torr. Two micro-channel plate (MCP) detectors upstream of the IC measured the local time-of-flight (TOF) of the recoils over a distance of 60 cm \cite{CV}. Recoils were then identified, and distinguished from ``leaky'' beam 
transmitted through the separator, by their locus on an energy loss-vs-TOF graph, an example of which is
shown in Figure \ref{etof}. Further discrimination was provided by prompt $\gamma$-rays detected in the BGO array in coincidence with events in the IC. The time between the prompt $\gamma$-ray detection and subsequent MCP detection allowed for a separator TOF measurement, which was used for additional particle identification. When the detection yields were too low to distinguish a clear $^{21}$Ne recoil locus, the profile likelihood technique \cite{Rolke} was used to calculate a confidence interval. In these instances, the MCP and separator TOF regions of interest were extrapolated from higher yield data.\\
\indent For each beam energy delivered, an energy measurement was made both with and without target gas present. In combination with the measurements of the gas target pressure, temperature and the known effective length \cite{Hutch2}, this allows the stopping power to be calculated. Beam energy measurement is performed by centering the beam on a set of slits at the energy-dispersed ion-optical focus after the first magnetic dipole field, using an NMR field read back, where the energy-to-field relationship for a given mass-to-charge ratio has been calibrated by many well-known, precise nuclear resonances \cite{Hutch,Hutch2}. The beam intensity was measured every hour in three Faraday cups (FC), one located upstream of   DRAGON, one after the gas target and one after the first dipole magnet. Continuous monitoring of the beam intensity throughout data taking was achieved via recoiling $\alpha$-particles, from elastic scattering of the beam on the helium in the target, detected in 
 two surface-barrier detectors located within the gas target assembly. These elastic scattering data were normalised to the measured beam intensity at the start and end of every run \cite{JDA}. Target pressures of between 4 and 8 Torr were used. The energy loss of the beam, in the center of mass, across the gas target varied from 53 keV at 8 Torr for the lowest energy, to 30 keV at 4 Torr for the highest energy.  For the five measurements around the $E_x$ = 8.155 MeV state, target pressures between 4 and 6 Torr were used, with a corresponding center of mass energy loss of 28 to 44 keV. 

At each energy, the raw yields were corrected for the separator efficiency, the charge state fraction for 4$^+$
recoils exiting the gas target, the effective efficiencies of the IC and
MCP detectors, and the data acquisition deadtime. As $\gamma$-ray coincidences were required for particle identification, the BGO array efficiency was also taken into account.
The separator efficiency was determined from Monte Carlo simulations of DRAGON using GEANT3 \cite{G3}. For centre of mass energies below $\approx$ 1 MeV, the maximum cone angle of the $^{21}$Ne recoil exceeds the DRAGON acceptance of 21 mrad. If a resonance is located upstream of the target centre, this limit is reached at higher energies. Similarly, the efficiency of the BGO array \cite{DG} depends on the location of the reaction in the target. For most energies studied in this work neither the width of the resonance, nor the angular distribution or decay scheme of the subsequent decay of $^{21}$Ne are known, and the measured statistics were too low to determine these values from the observed $\gamma$-ray energies and distributions. Simulations were, therefore, conducted assuming three decay schemes (direct to ground state, via the 3.74 MeV state, and via the 1.75 and 0.35 MeV states) and three angular distributions (isotropic, dipole, and quadrupole). For each simulated scenario (reaction location in gas target, assumed decay scheme, etc.) the corresponding separator transmission and BGO detection efficiencies were extracted, and the differences between the various scenarios used to determine the systematic errors on both values. \\
\indent Charge state distributions of $^{21}$Ne were measured at beam energies of 160, 202, 290, and 360 keV/u and the 4$^+$ charge state
fraction was estimated using an empirical formula from \cite{Wenjie}. This formula was used to interpolate the 4$^+$ charge state fraction for each of the recoil energies. The efficiency of the end detectors was taken from a comparison of MCP and IC event rate data using attenuated beam, together with the geometric transmission of the MCP detector grid.

\indent 
The effective cross sections ($\sigma$) and effective astrophysical S-factors ($S$) were then calculated from:
\begin{eqnarray}
\sigma = \frac{N_r}{N_b}\frac{A}{N_t} \\
S(E) = \frac{E}{e^{-2\pi\eta}}\sigma
\end{eqnarray}
 where $\frac{N_r}{N_b}$ is the corrected yield, $\frac{A}{N_t}$ is the reciprocal target nuclei per unit area, $e^{-2\pi\eta}$ is the Gamow factor and $E$ is the center of target center of mass energy. The resonance strength of the excitation level of interest was calculated via the equation \cite{Iliadis}

\begin{multline}
\omega \gamma = \frac{2\pi \epsilon(E_\mathrm{r}) Y}{\lambda ^2(E_\mathrm{r})}  \times \\ \left[\arctan\left(\frac{E_0-E_\mathrm{r}}{\Gamma /2}\right) - \arctan\left(\frac{E_0-E_\mathrm{r}-\Delta E}{\Gamma /2} \right) \right] ^{-1}
\label{eqn:yield}
\end{multline}  

\noindent where $\lambda$ is the system's de Broglie wavelength, $\epsilon$ is the target stopping power, $E_\mathrm{r}$ is the resonance energy, $E_0$ is the initial center of mass energy and $\Delta E$ is the beam energy loss across the entire length of the target. The target stopping power was calculated from
\begin{equation}
\epsilon(E) = -\frac{\mathrm{V}}{\mathrm{N}_\mathrm{t}}\frac{\mathrm{d}E}{\mathrm{d}x}
\end{equation}
where $\frac{\mathrm{V}}{\mathrm{N}_\mathrm{t}}$ is the reciprocal target density and $\frac{\mathrm{d}E}{\mathrm{d}x}$ is the rate of ion energy loss in the target. For the runs where the resonance was fully contained within the target, the thick target yield was used to calculate the resonance strength:
\begin{equation}
\omega\gamma = \frac{2\epsilon(E_r)}{\lambda^2(E_r)}Y_{max}.
\end{equation}

The stated errors include both systematic and statistical uncertainties. The main sources of systematic uncertainty were the BGO detection efficiency (10\%), separator transmission (between 20-30\% for the lower energy runs, 2-10\% for the 811 keV runs), detector efficiency and transmission (between 4-5\%) and integrated beam intensity (between 0.6-6\%). Uncertainties in stopping power (3.7\%) and recoil charge state fraction (1.6-4.1\%) were also accounted for. The range in uncertainies reflects the range of beam energies, populated states, recoil angular distribution and duration of the runs. 

\section{Results}

\begin{figure}[h!]
\centering\includegraphics[scale=0.45,angle=0]{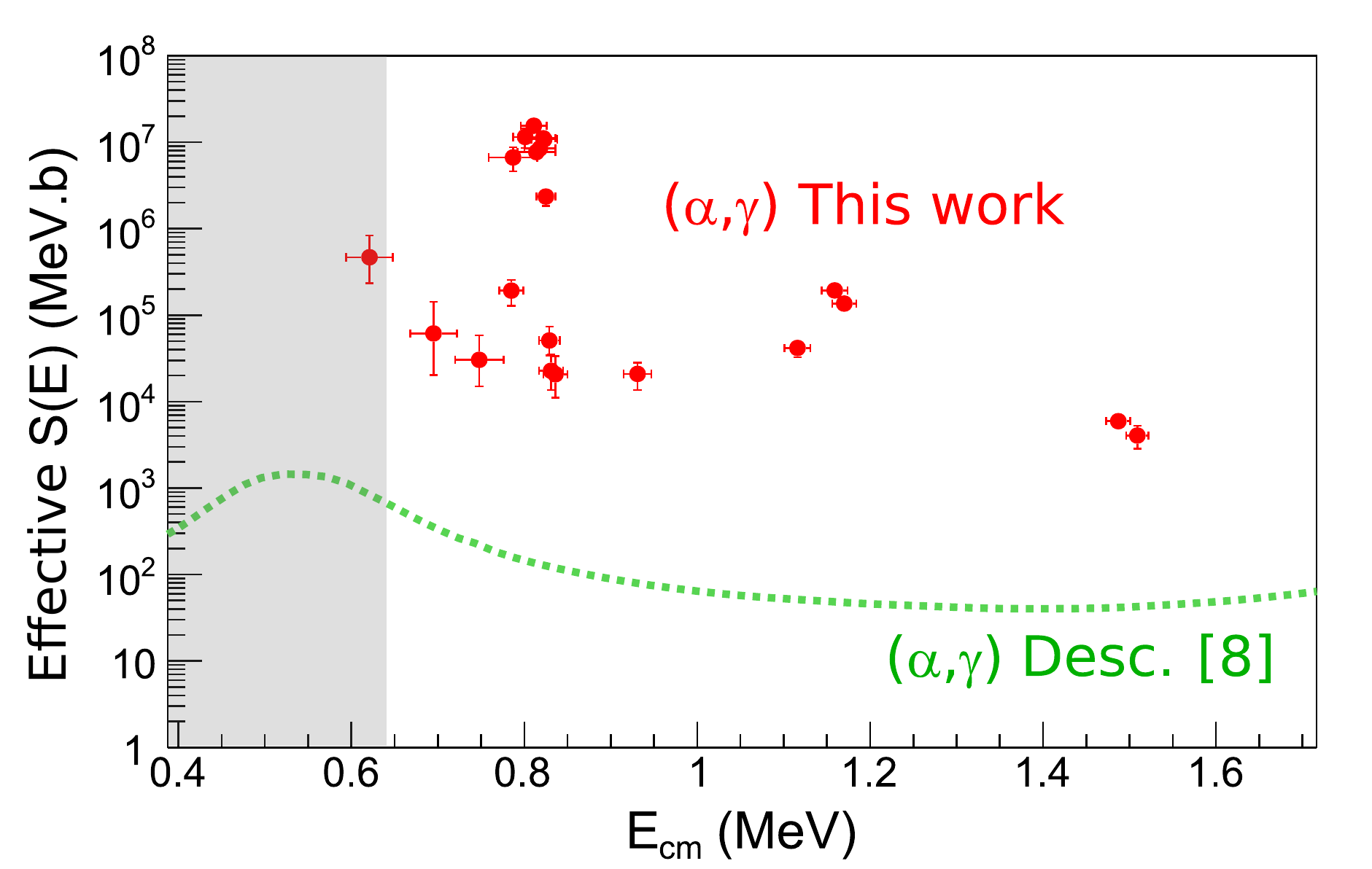}
\caption{\small (Color online) Effective astrophysical S-factor from the present work, compared to the calculation for the $^{17}$O($\alpha,\gamma$)$^{21}$Ne reaction from \cite{Desc}. Each data point represents the energy at the centre of the gas target and the horizontal error bar corresponds to the energy loss in the target. Target pressures of between 4 and 8 Torr were used.}
\label{Sfac1}
\end{figure}

Figure \ref{Sfac1} shows the measured S-factors at each centre of mass energy for the present work in comparison with the calculation from Descouvemont \cite{Desc}. It should be noted that the direct capture contribution is expected to be lower than the cross section from Descouvemont, and is thus considered negligible. Data were initially taken at several energies above 1 MeV, where the yield is much higher, to allow a comparison with the Descouvemont calculation. Measurements were then pushed lower towards the astrophysically interesting energy range. Table \ref{wgtab} gives the resonance strengths from the present work, compared to literature values where available. 

The data point around 1.1 MeV covers the state at 8.470 MeV.  A resonance strength of 1.9 $\pm$ 0.4 meV was determined, which is slightly higher than that reported by \cite{Best1} who found the strength to be 1.2 $\pm$ 0.2 meV.

The most prominent feature at around E$_{cm}$ = 0.81 MeV 
corresponds to a known J$^{\pi}$=(9/2)$^+$ state in $^{21}$Ne at an excitation energy of 8.155(1) MeV \cite{21Ne}. This resonance appears to be of comparable strength in both gamma and neutron channels \cite{Denker,Best2}. The weighted average of the five highest yield data points, where the resonance is fully within the gas target, gives a measured resonance strength of 5.4 $\pm$ 0.8 meV. This value is slightly weaker than the 7.6 $\pm$ 0.9 meV reported by \cite{Best1}. 

\indent 
There is a known J$^{\pi}$ = 3/2$^+$ state of total width 8 keV \cite{21Ne} at 8.069 MeV (E$_{cm}$ = 0.721 MeV). This state contributes to both the 0.695 and 0.748 MeV data points, with each measurement
covering approximately half of the relevant yield. This resonance has not previously been observed in ($\alpha,\gamma$)  and a strength of 8.7 $^{+7.0}_{-3.7}$ $\mu$eV was found. The quoted uncertainty does not include the uncertainty on the energy or the width of the state.

Between the measurement at 0.695 MeV and the lowest data point, there is a gap in the measured energy range, from 0.648 to 0.667 MeV, and so no constraint can be placed on the contribution of the 1/2$^-$ resonance at 0.66 MeV (E$_x$ = 8.009(10) MeV). However, as this state corresponds to an f-wave resonance and was observed as a neutron resonance, it is unlikely that this state will play any significant role in the ($\alpha,\gamma$) rate.\\ 
\indent The lowest data point measured lies inside the Gamow window for core helium burning. Three known states are covered by the energy thickness of the gas target at this beam energy (see Figure 1). Given the low yield, it is not possible to determine which state dominates and so a combined resonance strength of 4.0 $^{+3.1}_{-2.0}$ $\mu$eV is reported here. This value is a factor of around 10 lower than the 0.03-0.05 meV upper limit given in \cite{Best1}. 
The calculations by \cite{Best2} suggest that the 7.982 MeV level makes the dominant contribution here and so it is assumed that the observed strength comes from this state and a resonance energy of 0.633 MeV is used in the reaction rate calculation. However, if the full observed strength lies instead in the 0.612 MeV resonance, then the calculated reaction rate for the resonance would be 2.25 times higher.

\begin{table}
\begin{center}
\begin{tabular}{|c|c|c|}
\hline
E$_{CM}$   	&  		$\omega\gamma$  (meV) 	&	Literature value	\cite{Best1}				\\
(keV)    &      & (meV)   \\
\hline \hline
633 	  	&	     (4.0 $^{+3.1} _{-2.0}$ ) $\times$ 10$^{-3}$    	& -	\\
\hline
721       	&		  (8.7 $^{+7.0} _{-3.7}$ ) $\times$ 10$^{-3}$      &	 - 	\\
\hline
810  	  	&	5.4 $\pm$ 0.8  & 7.6 $\pm$ 0.9    \\ \hline
1122 & 1.9 $\pm$ 0.4 & 1.2 $\pm$ 0.2 \\
\hline
\end{tabular}
\end{center}
\caption[]{$^{17}$O($\alpha,\gamma$)$^{21}$Ne resonance strengths from the present work compared to literature values.} 
\label{wgtab}

\end{table}

\begin{figure}
\centering\includegraphics[scale=0.40,angle=0]{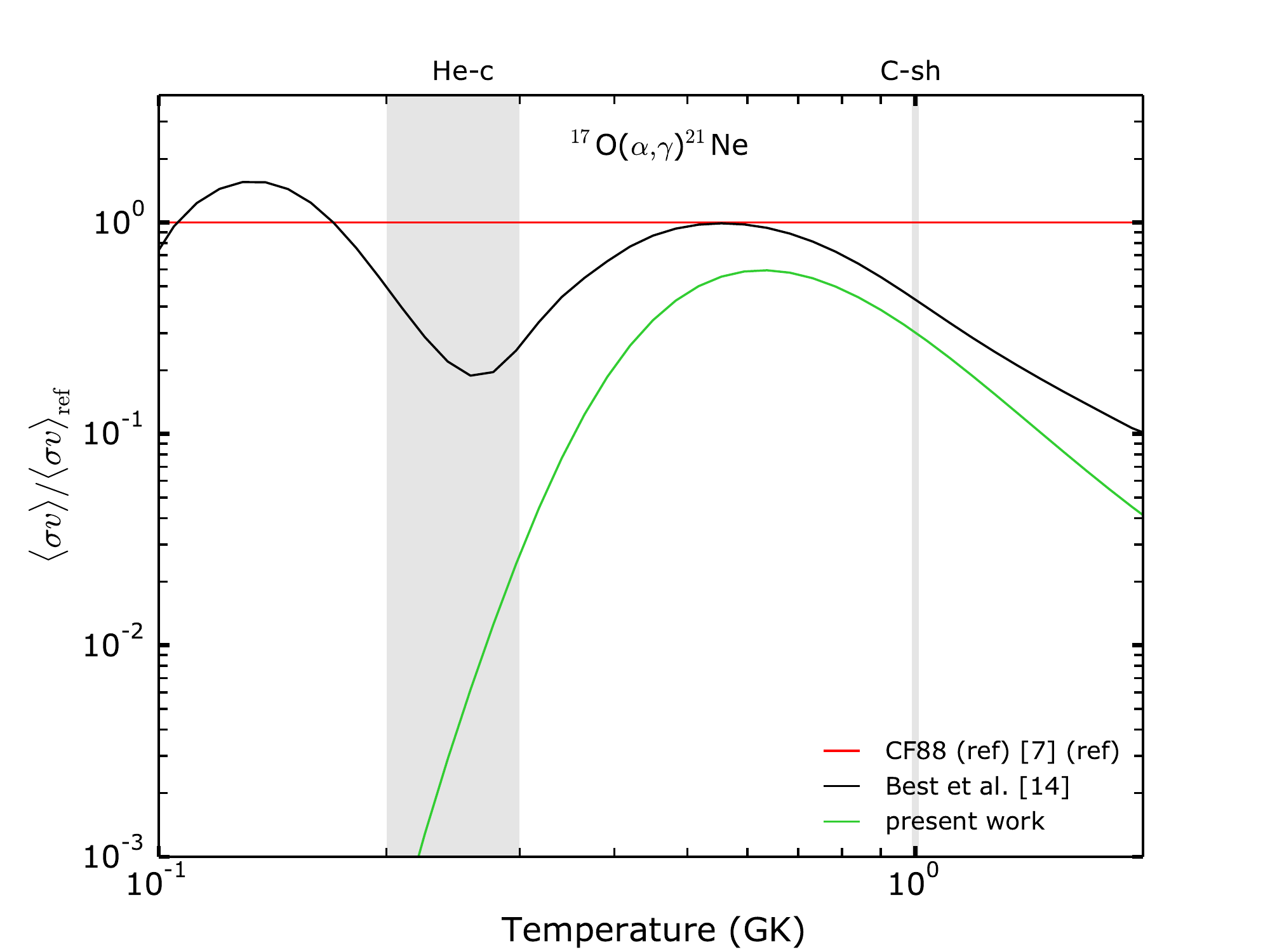}
\caption{\small (Color online) Ratio of $^{17}$O($\alpha,\gamma$)$^{21}$Ne reaction rates to CF88\cite{CF88}. The lowest curve (green) is from the present work and is a lower limit on the rate (see text). The upper curve (black) is the recommended rate of Best  {\it et al.} \cite{Best2}. The two shaded area indicate the approximate temperature in the helium burning core  and carbon burning shell of massive stars.}
\label{RR}
\end{figure}

\section{Astrophysical impact}
Using the narrow resonance formalism, the contributions to the reaction rate from the resonances at 0.633 and 0.81 MeV were calculated (the resonance at 0.721 MeV contributes less than 10\% to the total rate).  The sum of these two contributions (green) is shown in Figure \ref{RR}, in comparison with the recommended (black) rate from Best {\it et al.} \cite{Best2}, as a ratio to that of CF88. 
The cross section from the present work excludes the prediction of Descouvemont \cite{Desc}. However, the present rate is still around 100-1000 times lower than that of CF88 \cite{CF88}. It should be noted that within the Gamow window for helium core burning, there are 6 known states, giving a
typical level density of around 1.5 per 100 keV. This is well below that assumed for a statistical model approach and thus the Hauser-Feshbach treatment of this reaction at low energies used by CF88 \cite{CF88} may be expected to significantly overestimate the reaction rate. \\
\indent
It must be emphasised that the present rate should be considered as a lower limit. There are two 
known states in the energy region of interest whose spin and parity are not known, and none of the states below
E$_x$ = 7.96 MeV have experimentally constrained resonance strengths or partial widths.
Due to a lack of direct experimental data, the contribution of these states has not been included here. 
The recommended rate from Best {\it et al.} \cite{Best2} includes the contributions from 12 resonances which were not observed in that work, but whose resonance strengths have been calculated based on estimates of the 
$\alpha$-particle widths, branching between the $\gamma$- and neutron channels, and an assumed spectroscopic factor of 0.01.  It is therefore expected that the rate from the present work, based only on observed resonances, is significantly lower. However, within the Gamow window the difference between the present rate and the recommended rate from Best {\it et al.} \cite{Best2} is dominated by the estimated contribution from the resonance at 0.305 MeV. If this resonance is not as strong as suggested then the measured resonance at 0.633 MeV may make a significant contribution and the $^{17}$O($\alpha,\gamma$)$^{21}$Ne reaction rate would be closer to the lower limit presented here.

\indent
The reaction rate from the present work was tested in a 25 solar mass stellar model, at a metallicity of Z = 0.001 in mass fraction, and with an initial equatorial velocity of 70\% of the critical velocity (the velocity at which the gravitational force balances the centrifugal force). The model was computed with the Geneva stellar evolution code up to the core oxygen burning stage, with a network of 737 isotopes, fully coupled to the evolution (details can be found in \cite{Chop0} and \cite{Chop}). Figure 5 shows the yields of this model (green line) plus two additional models with the same ingredients except that one is computed with the recommended rate from Best {\it et al.} \cite{Best2} (black line) and the other with the recommended rate divided by 10 (red line). The latter rate was chosen to illustrate the impact of the 0.305 MeV resonance being weaker than estimated.

Significant differences are observed between yields from the present rate
and the recommended rate above strontium. These differences increase at higher atomic masses, with more than a factor of 10 around barium. The new rate leads to results closer to those using the recommended rate of Best {\it et al.} divided by a factor of 10 though the present rate leads to still higher production of elements around barium. 
It is clear that the current uncertainty in the $^{17}$O($\alpha,\gamma$)$^{21}$Ne reaction rate 
has a strong impact on the stellar model predictions. It is therefore crucial that, in the absence of direct measurements, the missing spectroscopic information (i.e. spin/parity, reduced energy uncertainty, partial widths) of the relevant states in $^{21}$Ne is determined to allow the reaction rate to be better constrained.

\begin{figure}
\centering\includegraphics[scale=0.30,angle=0]{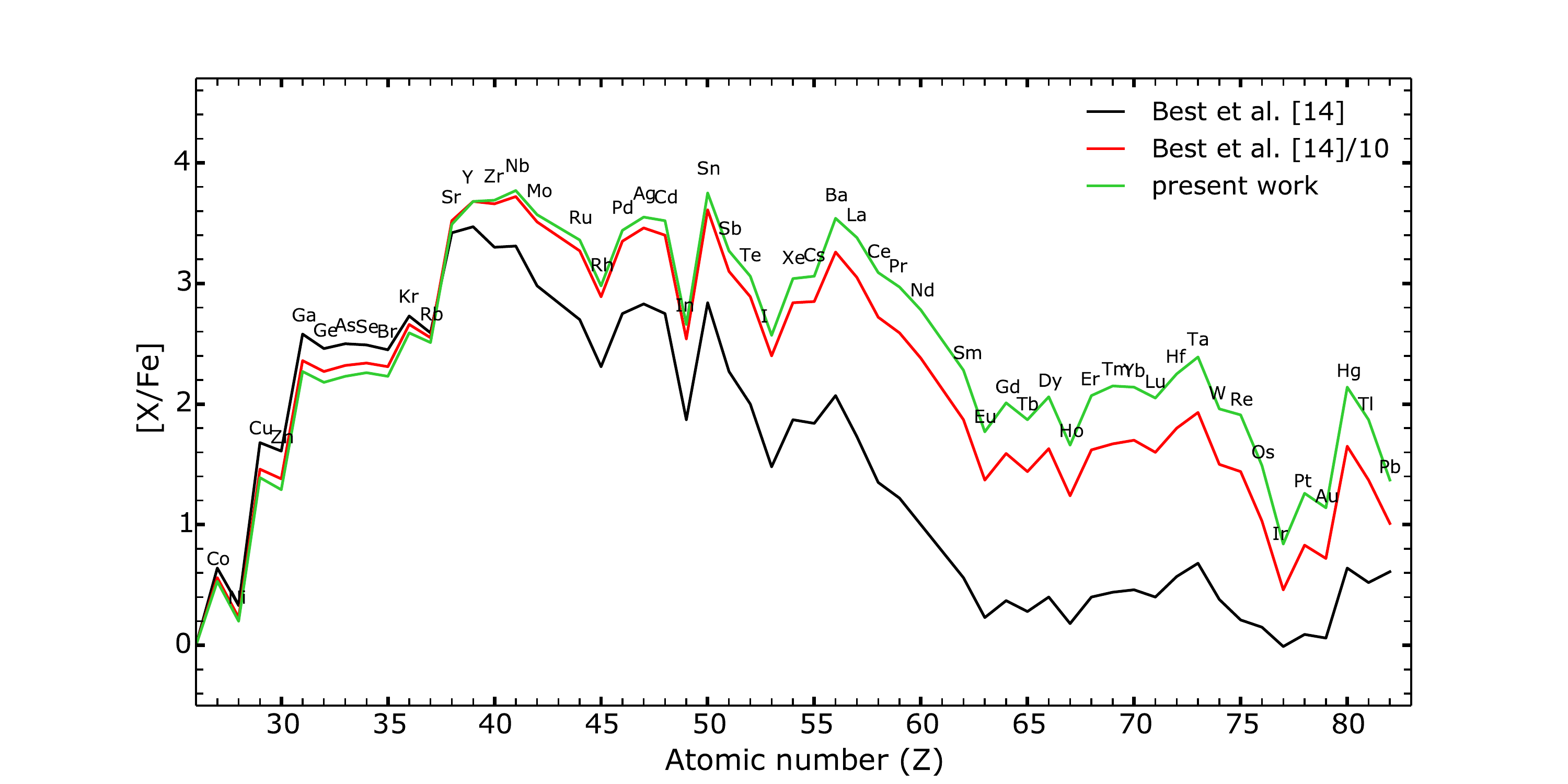}
\caption{\small (Color online) S-process yields of a fast rotating 25 M$_{\odot}$ at Z=0.001 when using the present rate, the recommended rate from \cite{Best2} and recommended rate/10 for the $^{17}$O($\alpha,\gamma$)$^{21}$Ne reaction (see text for further details). 
}
\label{yields}
\end{figure}

\section{Conclusions}
In conclusion, a direct measurement, in inverse kinematics, of the $^{17}$O($\alpha,\gamma$)$^{21}$Ne reaction has been performed at the DRAGON facility, at the TRIUMF
laboratory, Canada. Measurements were made of the reaction yield in the energy range E$_{cm}$ = 0.6 - 1.6 MeV, providing the only experimental data in the Gamow window for core helium burning. This work is over an order of magnitude more sensitive than previous work due to the enhanced discrimination provided by the coincident detection of both recoils and $\gamma$-rays. Moreover, the event identification does not require prior knowledge of the associated $\gamma$-ray energies. The abundances calculated with stellar models using the lower limit on the $^{17}$O($\alpha,\gamma$)$^{21}$Ne reaction rate from the present work show the maximum contribution to s-process production in low metallicity massive stars.


\section{Acknowledgments}
We would like to thank the beam delivery and ISAC operations groups at TRIUMF. In particular we gratefully acknowledge the invaluable assistance in beam production from K. Jayamanna, for delivering the high intensity beam. UK personnel were supported by the Science and Technology Facilities Council (STFC). Canadian authors were supported by the Natural Sciences and Engineering Research Council of Canada (NSERC). TRIUMF receives federal funding via a contribution agreement through the National Research Council of Canada. Authors acknowledge support from the ”ChETEC” COST Action (CA16117), supported by COST (European Cooperation in Science and Technology).
A. Choplin acknowledges funding from the Swiss National Science Foundation under grant P2GEP2-184492.  
RH acknowledges support from the World Premier International Research Center Initiative (WPI Initiative), MEXT, Japan.
The Colorado School of Mines group is supported via U.S. Department of Energy grant DE-FG02-93ER40789.
MP acknowledges support to NuGrid from NSF grant PHY-1430152
(JINA Center for the Evolution of the Elements) and STFC (through the University of Hull’s Consolidated Grant
ST/R000840/1), and access to viper, the University of Hull High Performance Computing Facility. MP acknowledges
the support from the ”Lendulet-2014” Programme of the Hungarian Academy of Sciences (Hungary). MP acknowledges support from the ERC Consolidator Grant (Hungary) funding scheme (project RADIOSTAR, G.A. n. 724560)

\end{document}